# $Q$-dependent Collective Relaxation Dynamics of Glass-Forming Liquid $Ca_{0.4}K_{0.6}(NO_3)_{1.4}$ Investigated by Wide-Angle Neutron Spin-Echo

Peng Luo[1], Yanqin Zhai[1,2], Peter Falus[3], Victoria García Sakai[4], Monika Hartl[5], Maiko Kofu[6], Kenji Nakajima[6], Antonio Faraone[7,*], Y Z[1,2,8,*]

[1]*Beckman Institute for Advanced Science and Technology, University of Illinois at Urbana-Champaign, Urbana, Illinois 61801, USA*

[2]*Department of Nuclear, Plasma, and Radiological Engineering, University of Illinois at Urbana-Champaign, Urbana, Illinois 61801, USA*

[3]*Institut Laue-Langevin (ILL), 38042 Grenoble, France*

[4]*ISIS Neutron and Muon Facility, Rutherford Appleton Laboratory, Science & Technology Facilities Council, Didcot OX11 0QX, UK*

[5]*European Spallation Source, SE-221 00 Lund, Sweden*

[6]*J-PARC Center, Japan Atomic Energy Agency, Tokai, Ibaraki 319-1195, Japan*

[7]*National Institute of Standards and Technology, 100 Bureau Drive, Gaithersburg, Maryland 20899-1070, USA*

[8]*Department of Electrical and Computer Engineering, University of Illinois at Urbana-Champaign, Urbana, Illinois 61801, USA*

*Email: antonio.faraone@nist.gov (A.F.); zhyang@illinois.edu (YZ)

## Abstract

Employing wide-angle neutron spin-echo spectroscopy, we measured the $Q$-dependent coherent intermediate scattering function of the prototypical ionic glass former $Ca_{0.4}K_{0.6}(NO_3)_{1.4}$, in the equilibrium and supercooled liquid states beyond the hydrodynamic regime. The data reveal a clear two-step relaxation: an exponential fast process, and a stretched exponential slow α-process. de Gennes narrowing is observed in all characteristic variables of the α-process: the relaxation time, amplitude, and stretching exponent. At all length scales probed, the relative amplitude of the α-relaxation decreases with increasing temperature and levels off in the normal liquid state. The temperature dependence of the stretching exponent and the relaxation time at different $Q$'s indicate that modifications of the relaxation mechanisms at the local length scales, manifested as temperature independent dynamic heterogeneity and smaller deviations from Arrhenius behavior, have occurred even above the α-β(Johari-Goldstein) bifurcation temperature.

Although glassy materials are ubiquitous in nature and various technologies, the understanding of glass formation remains a major scientific challenge, owing to the complex relaxation dynamics of supercooled liquids which exhibit extraordinary viscous slow-down upon cooling towards the glass transition [1]. Neutron spin-echo (NSE), which encodes the dynamic signal in the spin of the neutrons, offers the highest energy resolution among all neutron spectroscopy techniques, making it suitable for investigating the dynamics on pico- to nanosecond time scale [2]. NSE spectroscopy measures directly in time domain the intermediate scattering function (ISF), and has been extensively employed for experimental tests in the study of important theories of the glass transition, such as the mode coupling theory (MCT) [3–6]. NSE in combination with other quasi-elastic neutron scattering experiments have evidenced qualitatively some predictions of the MCT, but quantitative discrepancies between the theory and experimental observations for real glass-forming materials do exist [3–6]. To account for these discrepancies, it was proposed in the extended MCT that the atoms or molecules can still escape from cages by thermally activated hopping, at temperatures below the MCT predicted critical temperature $T_c$ [7,8]. Studies on typical polymeric and molecular glass formers employing NSE [9–11] and nuclear resonance X-ray scattering [12,13] have revealed, around $T_c$ and as the temperature is decreased, a transition from the slow α-relaxation to the Johari-Goldstein (JG) β-relaxation. This transition is not observed around the peak of the structure factor $S(Q)$, but in the wavevector transfer ($Q$) range of the valley, and is attributed to the local length scale of the JG β-process. Traditional NSE instruments only measure one $Q$ at a time, thus in previous NSE studies only the dynamics at few $Q$'s was investigated. An open question remains as to what the mechanism of the relaxations is at different length scales. To establish a detailed atomic-level description of the relaxation dynamics and understand the liquid-glass transition, direct determination of the $Q$-dependent ISF beyond the hydrodynamic regime is required [14].

To achieve this goal, we employed the new wide-angle spin-echo (WASP) instrument [15] at ILL, which employs anti-Helmholtz coils to allow simultaneous measurement of collective dynamics over a broad $Q$-range with unprecedented high data rate and accuracy [15]. Here, we scrutinized the relaxation dynamics of a prototypical fragile glass-forming liquid $Ca_{0.4}K_{0.6}(NO_3)_{1.4}$ (CKN, glass transition temperature $T_g \approx 333$ K and liquidus temperature $T_L \approx 483$ K [5]), in the equilibrium and supercooled liquid states, over the full range of the structure factor peak and the following valley, between 1.08 Å$^{-1}$ and 2.82 Å$^{-1}$. In this ionic system, coherent neutron scattering overwhelmingly dominates, therefore its relaxation dynamics has been extensively studied using neutron scattering [3,4,16–19]. However, due to the limited range of the $Q$-$t$-$T$ space and data accuracy probed by previous measurements, a full comparison with theories was not possible and some questions, such as whether the relaxation amplitude and the stretching exponent of the α-relaxation depend on temperature and $Q$ in the supercooled liquid, and how the relaxation dynamics are modified at different length scales, still remain unclear [16,17]. The data reported here with high accuracy and fully mapping a wide range of time, temperature and length scales, is used to address the controversies in previous studies and shed new light on the microscopic dynamics of glass forming liquids.

Wide-angle NSE measurements for CKN were carried out using the WASP spectrometer at temperatures between 383 K and 519 K. Additional NSE measurements extending the $Q$-range to the pre-peak of $S(Q)$ at 0.8 Å$^{-1}$ were performed using the NGA-NSE spectrometer [20] at NCNR. To access shorter times to complement the NSE data, Fourier deconvoluted time-of-flight spectra measured on the AMATERAS spectrometer [21,22] at J-PARC were compared with the ISFs collected on the WASP spectrometer. More details about the experimental methods are included in the Supplementary Materials [23].

Figure 1 shows representative ISFs, $I(Q,t)/I(Q,0)$, of CKN at 444 K for various $Q$'s [Fig. 1(a)], and at the structure factor maximum of 1.86 Å$^{-1}$ at various temperatures [Fig. 1(b)]. In comparison to previous data over the same time window [16,17], our data collected on the WASP spectrometer show much improved statistics and signal-to-noise ratio and reveal unambiguously the existence of a two-step relaxation. As one can see from Fig. 1(a), both the shape and amplitude of the slow α-process are highly dependent on $Q$, indicating length scale relevant relaxation dynamics. Figure 1(b) indicates that even in the equilibrium liquid state ($T > T_L$), the relaxation of CKN exhibits a well-defined two-step process.

In previous pioneering studies of CKN by Mezei [16,17], it has been shown that fits using a single stretched exponential function to account for a mix of fast process and slow α-process is invalid. An alternative procedure which eliminates the contribution of the fast process leads to more compatible results with the MCT, however, one has to assume *a priori*, that the amplitude of the slow α-process is temperature independent. Here, our data, which resolves a well-defined two-step dynamical process permits the fitting with a two-component model. An exponential function and a stretched exponential function, respectively, were used to describe the fast process and the slow α-process:

$$\frac{I(Q,t)}{I(Q,0)} = [1-f]\exp\left[-\left(\frac{t}{\tau_{\text{fast}}}\right)\right] + f\exp\left[-\left(\frac{t}{\tau_{\text{slow}}}\right)^{\beta}\right] \quad (1)$$

where $f$ is the amplitude of the slow α-relaxation (the effective Debye-Waller factor), $\tau_{\text{fast}}$ and $\tau_{\text{slow}}$ are the relaxation times of the fast process and the slow α-process, respectively, and $\beta$ the shape exponent of α-relaxation. Attempts to fit the data with only a stretched exponential function was not successful even for the highest temperature at 519 K.

The fitted $f$, $\tau_{\text{slow}}$, and $\beta$ are plotted against $Q$ in Fig. 2, which indicates that all parameters describing the density correlations are significantly $Q$-dependent and exhibit pronounced characteristic oscillations in phase with $S(Q)$. For low temperatures at 383 and 394 K, and at 1.08 Å$^{-1}$ of 402 K, the α-relaxation is outside the accessible time window of the measurement, therefore, the values of $\tau_{\text{slow}}$ and $\beta$ were excluded. As the initial plateau of the α-process is observed directly in these cases, the determination of $f$ is possible.

The α-relaxation time $\tau_{\text{slow}}$ follows an $S(Q)/Q^2$ dependence above 1.4 Å$^{-1}$ at all temperatures [Fig. 2(a)], in agreement with the observations in other glasses [6]. Such a behavior, known in general as de Gennes narrowing [24], indicates that the relaxation dynamics are significantly affected by the correlation between neighboring particles. Below 1.4 Å$^{-1}$, $\tau_{\text{slow}}$ is slightly larger than that predicted by the $S(Q)/Q^2$ dependence, likely related to the additional influence of medium range ordering as indicated by the pre-peak of $S(Q)$ around 0.8 Å$^{-1}$ [25].

Figure 2(b) shows that at the studied temperatures $\beta$ also varies in phase with $S(Q)$, between ≈0.45 and ≈0.75, being largest around the maximum of $S(Q)$. The same is observed for $f$ [Fig. 2(c)], which varies between ≈0.45 and ≈0.85 and shows a systematic variation in phase with $S(Q)$. The observation of the $Q$ dependence of $\beta$ is in agreement with the cases of *ortho*-terphenyl [6], isopropanol [26], hard-sphere system [27] and binary soft-sphere mixture [28], indicating again the important effect of particle-particle correlations on the collective relaxation dynamics. At the length scale specified by the structure factor peak, a narrower distribution of possible relaxation processes and thus larger $\beta$ is observed, reflecting the more coherent structure of the liquid at this $Q$. However, at $Q$'s outside this length scale, the structural correlation becomes weaker, leading to increased dynamic heterogeneity as manifested by the decreased value of $\beta$.

Figure 3 reports the temperature dependence of the fitting parameters, for the data measured on both the WASP spectrometer and the NGA-NSE spectrometer (Fig. S1). For clarity, the plot of $\beta$ as a function of temperature is divided into two panels as shown in Fig. 3(a). The left panel shows $\beta$ for $Q \leqslant 2.33$ Å$^{-1}$, where one can see that above 470 K $\beta$ is unaffected by temperature changes, while it decreases with decreasing temperature below 470 K; these findings indicate an increased dynamic heterogeneity of the α-relaxation in the supercooled liquid state [29]. This behavior is fully confirmed by the NSE measurements at NCNR, at 0.80 Å$^{-1}$ and 1.75 Å$^{-1}$ (half-filled triangles). As in the right panel we can see, for $Q$ = 2.49 Å$^{-1}$ $\beta$ also decreases with decreasing temperature below 470 K, but at a much slower rate than at lower $Q$'s. Eventually, for $Q \geqslant 2.60$ Å$^{-1}$, $\beta$ shows no systematic change with temperature.

Figure 3(b) is the Arrhenius plot of $\tau_{\text{slow}}$ fitted with the Vogel-Fulcher-Tammann (VFT) law [30],

$$\tau_{\text{slow}} = A \exp\left(\frac{DT_0}{T - T_0}\right) \tag{2}$$

where $D$ is a strength parameter that quantifies the deviations from Arrhenius behavior, with larger $D$ signifying smaller deviation from Arrhenius behavior. In Angell's fragility classification, CKN is a typical fragile glass-forming system [30]. In the studied temperature range, $\tau_{\text{slow}}$ can be well fitted by the VFT law [Fig. 3(b)]. The resulting $Q$-dependent $D$ values are plotted in the inset of Fig. 3(b) as red circles, together with a shaded magenta area that represents $D = 1.53\pm0.15$ obtained by VFT fit to the macroscopic shear relaxation time $\langle\tau\rangle = \eta/G_\infty$ in the temperature range of the present study [Fig. S2(a)], using the previously reported shear viscosity $\eta$ [31,32] and limiting high-frequency shear modulus $G_\infty$ [33]. We can see that below 2.4 Å$^{-1}$ $D$ remains basically unchanged between 1 and 2, in agreement with that of the macroscopic shear relaxation time, while above 2.4 Å$^{-1}$ it rapidly increases to ≈3 and further to ≈3.5 at larger $Q$'s. Note that below 2.4 Å$^{-1}$ the exponent $\beta$ shows rather weak or absent temperature dependence, in contrast to that at lower $Q$'s [Fig. 3(a)]. Therefore, to consider the influence of $\beta$, we calculated the mean relaxation time

$$\langle\tau_{\text{slow}}\rangle = \frac{\tau_{\text{slow}}}{\beta}\Gamma\left(\frac{1}{\beta}\right) \tag{3}$$

being $\Gamma(x)$ the gamma function, and fitted it with the VFT expression [Fig. S2(b)]. As the black squares indicate in the inset of Fig. 3(b), the influence of $\beta$ is negligible and the rapid increase of $D$ above 2.4 Å$^{-1}$ is retained for the case of $\langle\tau_{\text{slow}}\rangle$.

It has been revealed in polybutadiene and *ortho*-terphenyl that a transition from the α-process to the JG β-process occurs around $T_c$ in the $Q$ range above the primary peak of $S(Q)$, [10,12,13]. Whether this transition occurs in CKN is beyond the scope of the present study as we are at temperatures above $T_c$, which would be ≈368 K for CKN if it were to exist [4]. Intriguingly, our observations of contrasting behaviors of both $\beta$ and $D$ at the valley of $S(Q)$ ($Q > 2.4$ Å$^{-1}$) with respect to that around the peak indicate that even above $T_c$ the modifications of relaxation mechanisms at the local length scales are not trivial at all, at least for CKN. The VFT type temperature dependence of the relaxation time at all $Q$'s studied indicates that the slow α-relaxation dominates therein, while the increase of $D$, which means smaller deviations from Arrhenius behavior at $Q > 2.4$ Å$^{-1}$, suggests that the JG β-relaxation becomes dominant already above $T_c$ for the dynamics at the local length scales. Moreover, the rather slow decrease or even invariant exponent $\beta$ with decreasing temperature below $T_L$ for $Q > 2.4$ Å$^{-1}$, indicates that the dynamic heterogeneity at the local length scales is not affected by temperature, in contrast to the slow α-relaxation at $Q < 2.4$ Å$^{-1}$.

We note that this information was not available in previous NSE studies, where usually a $Q$-independent and/or $T$-independent $\beta$ had to be assumed *a priori* owing to the limited data accuracy even with an extended time window [4,9,10,16,17,34]. Furthermore, the previous NSE results of CKN were not able to resolve reliable value of $f$ for higher temperatures, e.g., at $T > 400$ K [4,16,17].

As reported in Fig. 3(c) we can see that at each $Q$ the effective Debye-Waller factor $f$ shows a linear decrease with increasing temperature until 470 K, above which it levels off. This observation is confirmed by the NSE measurements at NCNR at 0.80 Å$^{-1}$ and 1.75 Å$^{-1}$ (half-filled triangles), albeit with a small offset to the WASP data. Our data corroborate the previous result of Mezei et al. (open diamonds) [4] which found that the MCT prediction of a square-root singularity of the temperature dependence of $f$ never happens in CKN, and further demonstrate that a constant $f$ indeed exists but only in the normal liquid state. On the other hand, the marked temperature dependence of $f$ and $\beta$ in the supercooled liquid state contradicts the scaling (or time-temperature superposition) principle predicted by MCT for $T > T_c$, which was not established from the previous results of CKN [4,16,17].

The spin-echo measurements provide detailed information on the slow α-process, however, the fast process associated with the relaxation of particles before escaping from the cages formed by the nearest neighbors is not fully covered in the spin-echo time window, as shown in Fig. 1. To access short time scales less than 1 ps, we performed inelastic neutron scattering measurements for CKN on the AMATERAS spectrometer at J-PARC. The measured time-of-flight spectra were Fourier deconvoluted into the time domain ISF $I(Q,t)/I(Q,0)$. In Fig. 4 we plot representative data from the two measurements, the AMATERAS data at 420 and 480 K and the WASP data at 424 K and 483 K, at (1.26, 1.86 and 2.60) Å$^{-1}$, respectively. Note that the 3 K to 4 K temperature discrepancy between these two sets of data is trivial since the fast process is much less affected by temperature compared to the slow α-process. The match of the AMATERAS and the WASP data is unambiguously evident in the overlapping time range of 2.6 ps to 6 ps. We emphasize that we did not make any artificial adjustments to the data to force the match, in contrast to that done in previous studies [6,35,36]. Note that for time-of-flight measurements the neutron scattering signal is given by the sum of coherent ($I_{coh}$) and incoherent scattering ($I_{inc}$), by $I_{coh} + I_{inc}$, while for spin-echo measurements, the dynamic signal is encoded in the polarization of the beam and given by $I_{coh} - 1/3 I_{inc}$, where the incoherent scattering is highly suppressed. Therefore, the agreement between the AMATERAS and the WASP data also indicates that the incoherent scattering in CKN is negligible even at $Q$ values away from the peak of $S(Q)$. As seen in Fig. 4, these combined data can be well fitted to Eq. 1 with nearly the same parameters. The values of $\tau_{fast}$ from the fits are in the sub-picosecond time scale, which will be reported in detail elsewhere.

In summary, important insights of the collective relaxation dynamics in a glass-forming liquid CKN have been obtained by fully mapping the $Q$- and $T$-dependence of the coherent ISFs, only possible recently on the WASP neutron spectrometer at the ILL, France. Our study in the normal and supercooled liquid state of CKN, covering the full range of length scales across the structure factor peak and valley, reveals, unambiguously, a two-step relaxation. The fast exponential decay process is followed by a slow α-process of a stretched nature. The relaxation time, amplitude, and stretching exponent of the α-relaxation vary in phase with $S(Q)$, showing marked de Gennes narrowing. For all studied $Q$'s the amplitude of the α-relaxation decreases with increasing temperature and plateaus in the normal liquid state. For $Q < 2.4$ Å$^{-1}$ the stretching exponent increases with increasing temperature and levels off in the normal liquid state, however, for $Q > 2.4$ Å$^{-1}$ it increases rather slower and even becomes temperature independent at higher $Q$'s, accompanied by less deviations from Arrhenius behavior. These contrasting behaviors indicate that modifications of the relaxation mechanisms at the local length scales have already occurred even above the temperature where the α-relaxation transitions to the JG β-relaxation.

**ACKNOWLEDGMENTS**

We thank Tanya J. Dax, Yamali Hernandez, Donna A. Kalteyer and Michihiro Nagao for their help on the experiments. This work is supported by the U.S. Department of Energy, Office of Science, Office of Basic Energy Sciences, Materials Sciences and Engineering Division, under Award Number DE-SC0014084. The NSE measurements on the WASP spectrometer at ILL were performed under proposal number DIR-202, the data is accessible under doi:10.5291/ILL-DATA.DIR-202. Access to the NGA-NSE spectrometer at NCNR was provided by the Center for High Resolution Neutron Scattering, a partnership between the National Institute of Standards and Technology and the National Science Foundation under agreement No. DMR-1508249. The inelastic neutron scattering measurements on AMATERAS at J-PARC were performed based on the approved proposal No. 2019B0216.

**DISCLAIMER**

Certain trade names and company products are identified to specify adequately the experimental procedure. In no case does such identification imply our recommendation or endorsement, nor does it imply that the products are necessarily the best for the purpose. Throughout the paper, error bars of the raw data represent one standard deviation, and error bars of the fitted parameters represent one standard error of 95% confidence interval.

**References**

[1] P. G. Debenedetti and F. H. Stillinger, Nature **410**, 259 (2001).

[2] J. S. Gardner, G. Ehlers, A. Faraone, and V. García Sakai, Nat. Rev. Phys. **2**, 103 (2020).

[3] F. Mezei, W. Knaak, and B. Farago, Phys. Rev. Lett. **58**, 571 (1987).

[4] F. Mezei, W. Knaak, and B. Farago, Phys. Scr. **T19B**, 363 (1987).

[5] W. Petry and J. Wuttke, Transp. Theory Stat. Phys. **24**, 1075 (1995).

[6] A. Tölle, Reports Prog. Phys. **64**, 1473 (2001).

[7] W. Götze and L. Sjögren, J. Phys. C Solid State Phys. **21**, 3407 (1988).

[8] H. Z. Cummins, W. M. Du, M. Fuchs, W. Götze, S. Hildebrand, A. Latz, G. Li, and N. J. Tao, Phys. Rev. E **47**, 4223 (1993).

[9] D. Richter, B. Frick, and B. Farago, Phys. Rev. Lett. **61**, 2465 (1988).

[10] D. Richter, R. Zorn, B. Farago, B. Frick, and L. J. Fetters, Phys. Rev. Lett. **68**, 71 (1992).

[11] E. Rössler, Phys. Rev. Lett. **69**, 1620 (1992).

[12] M. Saito, S. Kitao, Y. Kobayashi, M. Kurokuzu, Y. Yoda, and M. Seto, Phys. Rev. Lett. **109**, 115705 (2012).

[13] T. Kanaya, R. Inoue, M. Saito, M. Seto, and Y. Yoda, J. Chem. Phys. **140**, 144906 (2014).

[14] T. Scopigno, G. Ruocco, and F. Sette, Rev. Mod. Phys. **77**, 881 (2005).

[15] P. Fouquet, G. Ehlers, B. Farago, C. Pappas, and F. Mezei, J. Neutron Res. **15**, 39 (2007).

[16] F. Mezei, Berichte Der Bunsengesellschaft Für Phys. Chemie **95**, 1118 (1991).

[17] F. Mezei, J. Non. Cryst. Solids **131**–**133**, 317 (1991).

[18] E. Kartini, M. Collins, and B. Collier, Phys. Rev. B **54**, 6292 (1996).

[19] M. Russina, F. Mezei, R. Lechner, S. Longeville, and B. Urban, Phys. Rev. Lett. **84**, 3630 (2000).

[20] N. Rosov, S. Rathgeber, and M. Monkenbusch, ACS Symp. Ser. **739**, 103 (1999).

[21] K. Nakajima, S. Ohira-Kawamura, T. Kikuchi, M. Nakamura, R. Kajimoto, Y. Inamura, N. Takahashi, K. Aizawa, K. Suzuya, K. Shibata, T. Nakatani, K. Soyama, R. Maruyama, H. Tanaka, W. Kambara, T. Iwahashi, Y. Itoh, T. Osakabe, S. Wakimoto, K. Kakurai, F. Maekawa, M. Harada, K. Oikawa, R. E. Lechner, F. Mezei, and M. Arm, J. Phys. Soc. Jpn. **80**, SB028 (2011).

[22] K. Nakajima, S. Ohira-Kawamura, M. Kofu, N. Murai, Y. Inamura, T. Kikuchi, and D. Wakai, JPS Conf. Proc. **33**, 011089 (2021).

[23] *See Supplemental Material* (n.d.).

[24] P. G. De Gennes, Physica **25**, 825 (1959).

[25] E. Kartini, M. F. Collins, B. Collier, F. Mezei, and E. C. Svensson, Can. J. Phys. **73**, 748 (1995).

[26] Y. Zhai, P. Luo, M. Nagao, K. Nakajima, T. Kikuchi, Y. Kawakita, P. A. Kienzle, Y. Z, and A. Faraone, Phys. Chem. Chem. Phys. **23**, 7220 (2021).

[27] M. Fuchs, I. Hofacker, and A. Latz, Phys. Rev. A **45**, 898 (1992).

[28] M. Fuchs and A. Latz, Physica A **201**, 1 (1993).

[29] R. Richert, J. Phys. Condens. Matter **14**, 201 (2002).

[30] R. Böhmer, K. L. Ngai, C. A. Angell, and D. J. Plazek, J. Chem. Phys. **99**, 4201 (1993).

[31] E. Rhodes, W. E. Smith, and A. R. Ubbelohde, Proc. R. Soc. London. Ser. A **285**, 263 (1965).

[32] R. Weiler, S. Blaser, and P. B. Macedo, J. Phys. Chem. **73**, 4147 (1969).

[33] L. M. Torell and R. Aronsson, J. Chem. Phys. **78**, 1121 (1983).

[34] A. Arbe, U. Buchenau, L. Willner, D. Richter, B. Farago, and J. Colmenero, Phys. Rev. Lett. **76**, 1872 (1996).

[35] M. Kiebel, E. Bartsch, O. Debus, F. Fujara, W. Petry, and H. Sillescu, Phys. Rev. B **45**, 10301 (1992).

[36] J. Wuttke, M. Kiebel, E. Bartsch, F. Fujara, W. Petry, and H. Sillescu, Zeitschrift Für Phys. B **91**, 357 (1993).

[37] Y. Inamura, T. Nakatani, J. Suzuki, and T. Otomo, J. Phys. Soc. Jpn **82**, SA031 (2013).

[38] R. T. Azuah, L. R. Kneller, Y. Qiu, P. L. W. Tregenna-Piggott, C. M. Brown, J. R. D. Copley, and R. M. Dimeo, J. Res. Natl. Inst. Stand. Technol. **114**, 341 (2009).

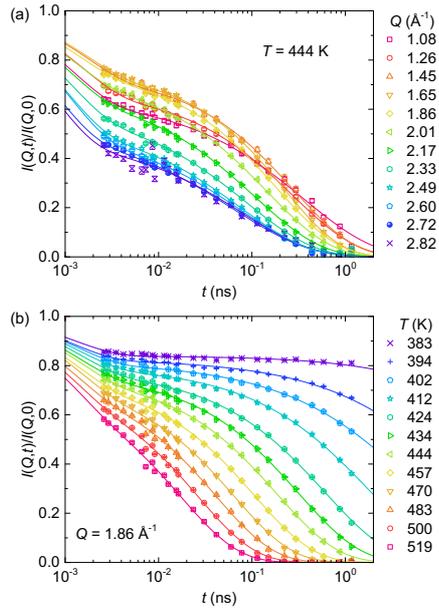

**Figure 1.** ISF of CKN measured at (a) 444 K and various $Q$'s between 1.08 Å$^{-1}$ and 2.82 Å$^{-1}$, and (b) 1.86 Å$^{-1}$ and various temperatures between 383 K and 519 K. The solid lines represent fit of Eq. 1.

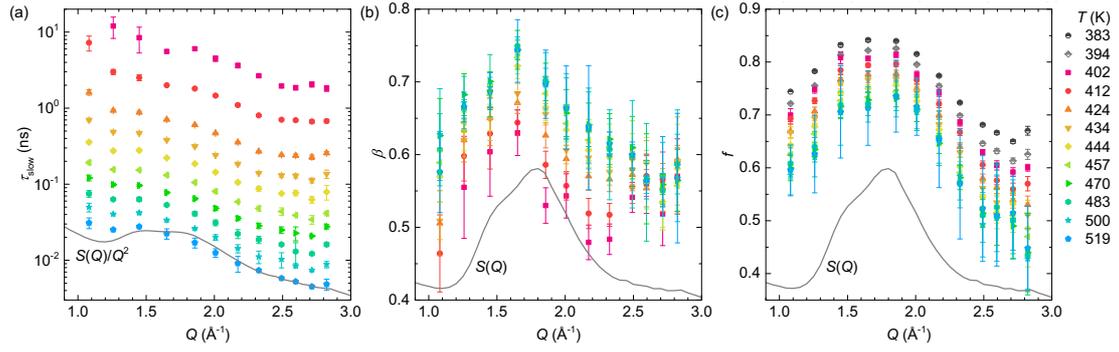

**Figure 2.** (a) $\tau_{\text{slow}}$, (b) $\beta$, and (c) $f$ as a function of $Q$ at various temperatures. The same legend applies to all three panels. The solid line in (a) represents $S(Q)/Q^2$ in logarithmic scale, and those in (b) and (c) are $S(Q)$.

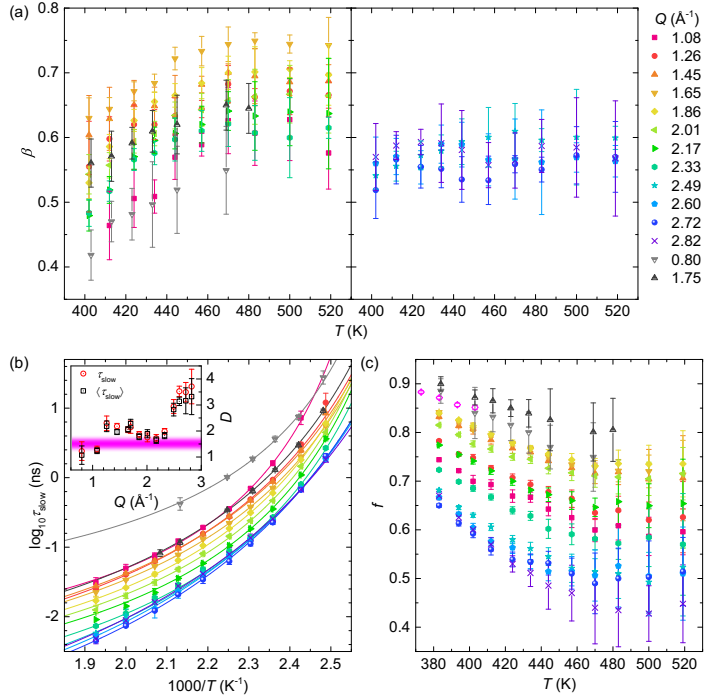

**Figure 3.** (a) $\beta$ as a function of temperature, (b) logarithmic $\tau_{\text{slow}}$ as a function of $1000/T$, (c) $f$ as a function of temperature, at various $Q$'s. The plot of $\beta$ is divided into two panels for clarity. The same legend applies to all panels. The half-filled triangles represent the results measured on NGA-NSE. The solid lines in (b) represent VFT fits (Eq. 2). The inset in (b) plots $D$ against $Q$, obtained by VFT fit to $\tau_{\text{slow}}$ (red circles) and $\langle\tau_{\text{slow}}\rangle$ (black squares), the shaded magenta area presents $D$ with its variation range obtained by VFT fit to the macroscopic shear relaxation time $\langle\tau\rangle$ in the temperature range of interest (for the VFT fit to $\langle\tau_{\text{slow}}\rangle$ and $\langle\tau\rangle$ see Fig. S2). The open diamonds in (c) are adapted from Ref. [4] at 1.86 Å$^{-1}$.

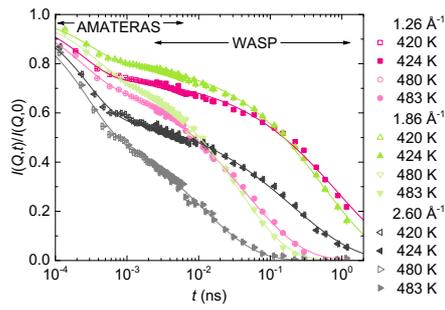

**Figure 4.** ISF of CKN covering a broad time window at three representative $Q$'s: (1.26, 1.86 and 2.60) Å$^{-1}$. The data presented by open symbols from 0.1 ps to 6 ps are Fourier deconvoluted time-of-flight spectra measured on AMATERAS, those presented by solid symbols from 2.6 ps to 1.2 ns are the WASP data. The AMATERAS data are measured at 420 K and 480 K, and the WASP data are at 424 K and 483 K. These two sets of data match in the overlapping time range without any artificial adjustment. The solid lines represent fit of Eq. 1 to the joint ISF.

Supplementary Materials for

**$Q$-dependent Collective Relaxation Dynamics of Glass-Forming Liquid Ca$_{0.4}$K$_{0.6}$(NO$_3$)$_{1.4}$ Investigated by Wide-Angle Neutron Spin-Echo**

Peng Luo, Yanqin Zhai, Peter Falus, Victoria García Sakai, Monika Hartl, Maiko Kofu, Kenji Nakajima, Antonio Faraone, Y Z

**The PDF file includes:**

Experimental Methods

Figure S1

Figure S2

**Experimental Methods**

CKN is a hydroscopic material, therefore, much care should be taken in the preparation process. The CKN sample was obtained by mixing appropriate weight of high-purity calcium nitrate tetrahydrate (Ca(NO$_3$)$_2$•4H$_2$O, 99.95% purity) and potassium nitrate (KNO$_3$, 99.995% purity) purchased from Sigma-Aldrich; the mixture was heated slowly up to above the melting point and dried in a vacuum oven at 523 K for over 48 hours. The dehydrated mixture was then transferred immediately to a glove box and heated again to 523 K for 12 hours under vacuum to remove any possible water absorption from the atmosphere that may have occurred during the transfer. Then an appropriate amount (≈10 g) of molten mixture was sealed into a vanadium annular can (12.7 mm inner diameter and 1.2 mm thickness) in the glove box filled with high-purity helium.

Wide-angle NSE measurements were performed using the high-intensity WASP spectrometer [15] at the Institut Laue-Langevin (ILL), France, which enables high-precise measurements for multiple wavevector transfer ($Q$) at the same time. To enable access to the $Q$ range from 1.08 to 2.82 Å$^{-1}$ which covers the structure factor peak (at 1.86 Å$^{-1}$) and valley of CKN, the incoming neutron wavelength λ was set to 4 Å. The data were collected at temperatures between 383 K and 519 K. The temperature was controlled with a closed cycle refrigerator. Before measurement at each temperature, the sample was first heated up to 530 K and equilibrated for 30 minutes, then directly cooled in an identical fashion each time to the target temperature and equilibrated for another 30 minutes, allowing sample equilibration and well-defined thermal history for each run. We run diffraction after each measurement to ensure that no crystallization occurred in the sample. Additional NSE measurements were performed using the NGA-NSE spectrometer at the National Institute of Standards and Technology (NIST) Center for Neutron Research (NCNR) [20]. An incoming neutron wavelength of 5 Å was used and the measurements were performed for two $Q$'s, 0.8 Å$^{-1}$ and 1.75 Å$^{-1}$, corresponding to the pre-peak and the main peak of $S(Q)$.

Inelastic neutron scattering measurements were carried out on the cold-neutron disk-chopper spectrometer AMATERAS [21,22] at J-PARC in Japan, using a neutron wavelength of 3.26 Å. The time of flight data of AMATERAS are processed by using the software suite Utsusemi [37]. A measurement of the empty vanadium can at 300 K was used as the instrument energy resolution for the data deconvolution. Normalized Fourier deconvoluted time-of-flight spectra were compared with the intermediate scattering function collected on the WASP spectrometer. The structure factor $S(Q)$ was obtained by integration at [−0.1, 0.1] meV around the elastic channel.

For the WASP data reduction an ILL developed software based on the commercial software package Igor Pro 8 was used [15]. The software DAVE [38] was used for the NGA-NSE data reduction and the Fourier deconvolution of the AMATERAS data. The transmissions of the sample and the empty can were measured at 100 K on the NGA-NSE spectrometer at NCNR using 5 Å neutron beam, which indicates a self-shielding factor of ≈85%, hence we estimate that double scattering would account for at most ≈15% of the measured scattering at some $Q$'s and therefore multiple scattering corrections have been deemed unnecessary.

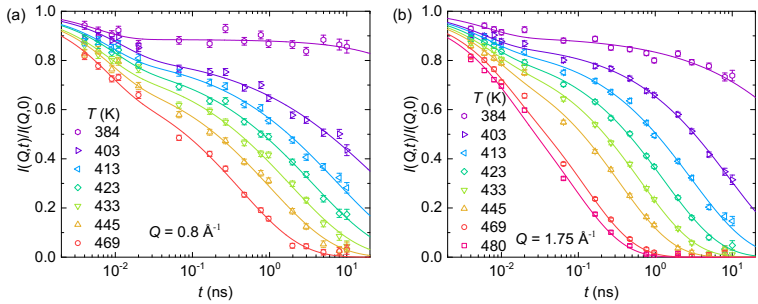

**Figure S1.** (a) Intermediate scattering function of CKN as measured on the NGA-NSE spectrometer at NCNR for (a) $Q = 0.8$ Å$^{-1}$ and (b) $Q = 1.75$ Å$^{-1}$ at various temperatures. The solid lines represent fit of Eq. 1.

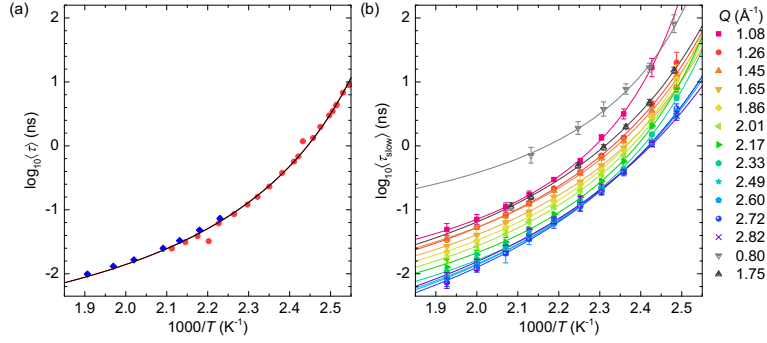

**Figure S2.** (a) The macroscopic shear relaxation time ⟨τ⟩, and (b) the average relaxation time ⟨$τ_{slow}$⟩ at various $Q$'s, as a function of inverse of temperature 1000/$T$. ⟨τ⟩ was calculated with the shear viscosity $η$ (blue diamonds from Ref. [31], red circles from Ref. [32]) and the limiting high-frequency shear modulus $G_∞$ [33] by ⟨τ⟩ = $η/G_∞$. The half-filled triangles are fitting results of the data measured on the NGA-NSE spectrometer at NCNR shown in Fig. S1.

**DISCLAIMER**

**References**

[1] P. G. Debenedetti and F. H. Stillinger, Nature **410**, 259 (2001).

[2] J. S. Gardner, G. Ehlers, A. Faraone, and V. García Sakai, Nat. Rev. Phys. **2**, 103 (2020).

[3] F. Mezei, W. Knaak, and B. Farago, Phys. Rev. Lett. **58**, 571 (1987).

[4] F. Mezei, W. Knaak, and B. Farago, Phys. Scr. **T19B**, 363 (1987).

[5] W. Petry and J. Wuttke, Transp. Theory Stat. Phys. **24**, 1075 (1995).

[6] A. Tölle, Reports Prog. Phys. **64**, 1473 (2001).

[7] W. Götze and L. Sjögren, J. Phys. C Solid State Phys. **21**, 3407 (1988).

[8] H. Z. Cummins, W. M. Du, M. Fuchs, W. Götze, S. Hildebrand, A. Latz, G. Li, and N. J. Tao, Phys. Rev. E **47**, 4223 (1993).

[9] D. Richter, B. Frick, and B. Farago, Phys. Rev. Lett. **61**, 2465 (1988).

[10] D. Richter, R. Zorn, B. Farago, B. Frick, and L. J. Fetters, Phys. Rev. Lett. **68**, 71 (1992).

[11] E. Rössler, Phys. Rev. Lett. **69**, 1620 (1992).

[12] M. Saito, S. Kitao, Y. Kobayashi, M. Kurokuzu, Y. Yoda, and M. Seto, Phys. Rev. Lett. **109**, 115705 (2012).

[13] T. Kanaya, R. Inoue, M. Saito, M. Seto, and Y. Yoda, J. Chem. Phys. **140**, 144906 (2014).

[14] T. Scopigno, G. Ruocco, and F. Sette, Rev. Mod. Phys. **77**, 881 (2005).

[15] P. Fouquet, G. Ehlers, B. Farago, C. Pappas, and F. Mezei, J. Neutron Res. **15**, 39 (2007).

[16] F. Mezei, Berichte Der Bunsengesellschaft Für Phys. Chemie **95**, 1118 (1991).

[17] F. Mezei, J. Non. Cryst. Solids **131**–**133**, 317 (1991).

[18] E. Kartini, M. Collins, and B. Collier, Phys. Rev. B **54**, 6292 (1996).

[19] M. Russina, F. Mezei, R. Lechner, S. Longeville, and B. Urban, Phys. Rev. Lett. **84**, 3630 (2000).

[20] N. Rosov, S. Rathgeber, and M. Monkenbusch, ACS Symp. Ser. **739**, 103 (1999).

[21] K. Nakajima, S. Ohira-Kawamura, T. Kikuchi, M. Nakamura, R. Kajimoto, Y. Inamura, N. Takahashi, K. Aizawa, K. Suzuya, K. Shibata, T. Nakatani, K. Soyama, R. Maruyama, H. Tanaka, W. Kambara, T. Iwahashi, Y. Itoh, T. Osakabe, S. Wakimoto, K. Kakurai, F. Maekawa, M. Harada, K. Oikawa, R. E. Lechner, F. Mezei, and M. Arm, J. Phys. Soc. Jpn. **80**, SB028 (2011).

[22] K. Nakajima, S. Ohira-Kawamura, M. Kofu, N. Murai, Y. Inamura, T. Kikuchi, and D. Wakai, JPS Conf. Proc. **33**, 011089 (2021).

[23] *See Supplemental Material* (n.d.).

[24] P. G. De Gennes, Physica **25**, 825 (1959).

[25] E. Kartini, M. F. Collins, B. Collier, F. Mezei, and E. C. Svensson, Can. J. Phys.

**73**, 748 (1995).

[26] Y. Zhai, P. Luo, M. Nagao, K. Nakajima, T. Kikuchi, Y. Kawakita, P. A. Kienzle, Y. Z, and A. Faraone, Phys. Chem. Chem. Phys. **23**, 7220 (2021).

[27] M. Fuchs, I. Hofacker, and A. Latz, Phys. Rev. A **45**, 898 (1992).

[28] M. Fuchs and A. Latz, Physica A **201**, 1 (1993).

[29] R. Richert, J. Phys. Condens. Matter **14**, 201 (2002).

[30] R. Böhmer, K. L. Ngai, C. A. Angell, and D. J. Plazek, J. Chem. Phys. **99**, 4201 (1993).

[31] E. Rhodes, W. E. Smith, and A. R. Ubbelohde, Proc. R. Soc. London. Ser. A **285**, 263 (1965).

[32] R. Weiler, S. Blaser, and P. B. Macedo, J. Phys. Chem. **73**, 4147 (1969).

[33] L. M. Torell and R. Aronsson, J. Chem. Phys. **78**, 1121 (1983).

[34] A. Arbe, U. Buchenau, L. Willner, D. Richter, B. Farago, and J. Colmenero, Phys. Rev. Lett. **76**, 1872 (1996).

[35] M. Kiebel, E. Bartsch, O. Debus, F. Fujara, W. Petry, and H. Sillescu, Phys. Rev. B **45**, 10301 (1992).

[36] J. Wuttke, M. Kiebel, E. Bartsch, F. Fujara, W. Petry, and H. Sillescu, Zeitschrift Für Phys. B **91**, 357 (1993).

[37] Y. Inamura, T. Nakatani, J. Suzuki, and T. Otomo, J. Phys. Soc. Jpn **82**, SA031 (2013).

[38] R. T. Azuah, L. R. Kneller, Y. Qiu, P. L. W. Tregenna-Piggott, C. M. Brown, J. R. D. Copley, and R. M. Dimeo, J. Res. Natl. Inst. Stand. Technol. **114**, 341 (2009).